\documentclass{JHEP3} 

\JHEPspecialurl{http://jhep.sissa.it/JOURNAL/JHEP3.tar.gz}

\usepackage{epsfig,multicol,bbm,cite}

\newcommand\fverb{\setbox\pippobox=\hbox\bgroup\verb}
\newcommand\fverbdo{\egroup\medskip\noindent%
			\fbox{\unhbox\pippobox}\ }
\newcommand\fverbit{\egroup\item[\fbox{\unhbox\pippobox}]}
\newbox\pippobox

\newcommand{\Eq}[1]{Eq.~(\ref{#1})}
\newcommand{\Eqs}[1]{Eqs.~(\ref{#1})}
\newcommand{\andEq}[1]{and~(\ref{#1})}
\newcommand{\Fig}[1]{Fig.~\ref{#1}}
\newcommand{\Ref}[1]{Ref.~\cite{#1}}

\newcommand{\Tab}[1]{Tab.~\ref{#1}}

\newcommand{\ie}{{\em i.e.}}

\newcommand{\VA}[3]{\ifthenelse{\equal{#2}{#3}}
{\ensuremath{#1\pm#2}}{\ensuremath{#1\,^{+#2}_{-#3}}}}

\newcommand{\pvec}{\ensuremath{\mathbf{p}}}
\newcommand{\rxy}{\ensuremath{r_{xy}}}

\newcommand{\xvec}{\ensuremath{\mathbf{x}}}

\newcommand{\DKeIII}{\ensuremath{K_{e3}}}

\newcommand{\DKmuIII}{\ensuremath{K_{\mu 3}}}

\def \gev  {{\rm \,Ge\kern-0.125em V}}
\def \mev  {{\rm \,Me\kern-0.125em V}}
\def \kev  {{\rm \,ke\kern-0.125em V}}
\def \ev   {{\rm \,e\kern-0.125em V}}

\newcommand{\dTCA}{\ensuremath{d_{\mathrm{TC}}}}
\newcommand{\dtTCA}{\ensuremath{d_{\perp,\,\mathrm{TC}}}}

\newcommand{\xc}{\ensuremath{\mathbf{x}_\mathrm{c}}}
\newcommand{\pc}{\ensuremath{\mathbf{p}_\mathrm{c}}}
\newcommand{\lc}{\ensuremath{l_\mathrm{c}}}
\newcommand{\dc}{\ensuremath{d_\mathrm{c}}}

\newcommand{\klpln}{\mbox{$K_{L}\to\pi^{\pm}\ell^{\mp}\nu$}}
\newcommand{\klpmn}{\mbox{$K_{L}\to\pi\mu\nu$}}
\newcommand{\semil}{\mbox{$K_{\mu3}$}}

\newcommand{\fphat}{\mbox{$\tilde{f}_+(t)$}}

\newcommand{\fzhat}{\mbox{$\tilde{f}_0(t)$}}

\newcommand{\miss}{\mbox{$E_{\rm miss}-p_{\rm miss}$}}

\newcommand{\kspp}{\mbox{$K_{S}\to\pi^+\pi^- $}}

\newcommand{\klpen}{\mbox{$K_{L}\to\pi^{\pm}e^{\mp}\nu$}}

\def\ifm#1{\relax\ifmmode#1\else$#1$\fi} \def\DAF{DA\char8NE} \def\x{\ifm{\times}}
\def\pt#1,#2,{\ifm{#1\x10^{#2}}} \def\up#1{\ifm{^{#1}}} 
\def\ab{\ifm{\sim}}   \def\to{\ifm{\rightarrow}}
\def\kl{\ifm{K_L}} \def\ks{\ifm{K_S}}  
  
\def\po{\ifm{\pi^0}} \def\pic{\ifm{\pi^+\pi^-}}    
\def\rmk{\rm\kern.5mm }   \def\dif{\hbox{d}} \def\f{\ifm{\phi}}  \def\minus{$-$}
 \def\bye{\end{document}}

\catcode`@=11 
\newdimen\z@ \z@=0pt 
\newskip\z@skip \z@skip=0pt plus0pt minus0pt
\def\m@th{\mathsurround=\z@}
\def\ialign{\everycr{}\tabskip\z@skip\halign} 
\def\eqalign#1{\null\,\vcenter{\openup\jot\m@th
  \ialign{\strut\hfil$\displaystyle{##}$&$\displaystyle{{}##}$\hfil
      \crcr#1\crcr}}\,}
\catcode`@=12 
\def\figb#1;#2;{\parbox{#2cm}{\epsfig{file=#1.eps,width=#2cm}}}
\let\cl=\centerline
\def\figbc#1;#2;{\cl{\figb #1;#2;}}

\newcommand{\lam}{\mbox{$\lambda_+$}}
\newcommand{\lamp}{\mbox{$\lambda'_+$}}
\newcommand{\lampp}{\mbox{$\lambda''_+$}}
\newcommand{\lamo}{\mbox{$\lambda_0$}}
\newcommand{\lamop}{\mbox{$\lambda'_0$}}
\newcommand{\lamopp}{\mbox{$\lambda''_0$}}

\def\dlp{\hbox{$\delta\lambda_+'$}}
\def\dlpp{\hbox{$\delta\lambda_+''$}} \def\dop{\hbox{$\delta\lambda_0'$}}
\def\dopp{\hbox{$\delta\lambda_0''$}}

\title{\mathversion{bold}Measurement of the \kl\ \to\ $\pi\mu\nu$ form factor parameters with the KLOE detector}
\author{The KLOE collaboration:\\
F.~Ambrosino,$^d$
A.~Antonelli,$^a$
M.~Antonelli,$^a$
F.~Archilli,$^a$
C.~Bacci,$^g$
P.~Beltrame,$^d$
G.~Bencivenni,$^a$
S.~Bertolucci,$^a$
C.~Bini,$^e$
C.~Bloise,$^a$
S.~Bocchetta,$^g$
F.~Bossi,$^a$
P.~Branchini,$^g$
R.~Caloi,$^f$
P.~Campana,$^a$
G.~Capon,$^a$
T.~Capussela,$^a$
F.~Ceradini,$^g$
S.~Chi,$^a$
G.~Chiefari,$^d$
P.~Ciambrone,$^a$
E.~De~Lucia,$^a$
A.~De~Santis,$^f$
P.~De~Simone,$^a$
G.~De~Zorzi,$^f$
A.~Denig,$^b$
A.~Di~Domenico,$^f$
C.~Di~Donato,$^d$
B.~Di~Micco,$^g$
A.~Doria,$^d$
M.~Dreucci,$^a$
G.~Felici,$^a$
A.~Ferrari,$^a$
M.~L.~Ferrer,$^a$
S.~Fiore,$^f$
C.~Forti,$^a$
P.~Franzini,$^f$
C.~Gatti,$^a$
P.~Gauzzi,$^f$
S.~Giovannella,$^a$
E.~Gorini,$^c$
E.~Graziani,$^g$
W.~Kluge,$^b$
V.~Kulikov,$^j$
F.~Lacava,$^f$
G.~Lanfranchi,$^a$
J.~Lee-Franzini,$^{a,h}$
D.~Leone,$^b$
M.~Martini,$^a$
P.~Massarotti,$^d$
W.~Mei,$^a$
S.~Meola,$^d$
S.~Miscetti,$^a$
M.~Moulson,$^a$
S.~M\"uller,$^a$
F.~Murtas,$^a$
M.~Napolitano,$^d$
F.~Nguyen,$^g$
M.~Palutan,$^a$
E.~Pasqualucci,$^f$
A.~Passeri,$^g$
V.~Patera,$^{a,e}$
F.~Perfetto,$^d$
M.~Primavera,$^c$
P.~Santangelo,$^a$
G.~Saracino,$^d$
B.~Sciascia,$^a$
A.~Sciubba,$^{a,e}$
A.~Sibidanov,$^{a}$
T.~Spadaro,$^a$
M.~Testa,$^f$
L.~Tortora,$^g$
P.~Valente,$^f$
G.~Venanzoni,$^a$
R.Versaci,$^a$
G.~Xu,$^{a,i}$
\\
\llap{$^a$}Laboratori Nazionali di Frascati dell'INFN, Frascati, Italy\\
\llap{$^b$}Institut f\"ur Experimentelle Kernphysik, Universit\"at Karlsruhe, Germany\\
\llap{$^c$}Dipartimento di Fisica dell'Universit\`a e Sezione INFN, Lecce, Italy\\
\llap{$^d$}Dipartimento di Scienze Fisiche dell'Universit\`a  ``Federico II'' e Sezione INFN, Napoli, Italy\\
\llap{$^e$}Dipartimento di Energetica dell'Universit\`a ``La Sapienza'', Roma, Italy\\
\llap{$^f$}Dipartimento di Fisica dell'Universit\`a ``La Sapienza'' e Sezione INFN, Roma, Italy\\
\llap{$^g$}Dipartimento di Fisica dell'Universit\`a ``Roma Tre'' e Sezione INFN, Roma, Italy\\
\llap{$^h$}Physics Department, State University of New York at Stony Brook, USA\\
\llap{$^i$}Institute of High Energy Physics of Academia Sinica,  Beijing, China\\
\llap{$^j$}Institute for Theoretical and Experimental Physics, Moscow, Russia\\
}
\received{\today} 		
\revised{\today}
\accepted{\today}		

\abstract{
 Using 328 pb\up{-1}of data collected at \DAF\ corresponding to \ab1.8 million \klpmn\ decays, we have
 measured the \semil\ form factor parameters. The structure of the $K-\pi$ vector-current provides
 information about the dynamics of the strong interaction; its knowledge is necessary for evaluation
 of the phase-space integral required for measuring the CKM matrix element $V_{us}$
 and for testing lepton universality in kaon decays.
 Using a new parametrization for the vector and scalar form factors, we find
 $\lambda_+$=\pt(25.7\pm 0.6),-3, and  $\lambda_0$=\pt(14.0\pm 2.1),-3,.
Our result for $\lambda_0$, together with recent lattice calculations of $f_\pi$, $f_K$ and $f(0)$, satisfies the Callan-Treiman relation.
}


\begin{document}
\section{Introduction}
\FIGURE{\parbox{6cm}{\figbc pictures/kel3ffgray;5;}
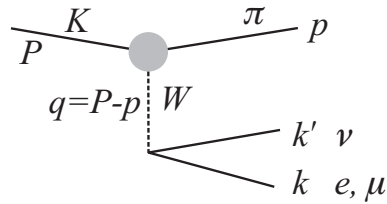
\caption{Amplitude for \klpln. The gray circle indicates the $K\pi W$ vertex structure.}\label{fig:diag}}%
 Semileptonic kaon decays, \klpln,\break (\Fig{fig:diag}) offer possibly the cleanest way to obtain an
 accurate value of the Cabibbo angle, or better, $V_{us}$.
Since $K\to \pi$ is a $0^-\to 0^-$  transition, only the vector part of the hadronic weak current
 has a non-vanishing contribution.
 Vector transitions are protected by the Ademollo-Gatto theorem against SU(3) breaking corrections to
 lowest order in $m_s$ (or $m_s-m_{u,d}$). At present, the largest uncertainty in calculating $V_{us}$ from
 the decay rate is due to the difficulties in  computing the matrix element
 $\langle\pi|J^{\rm had}_{\alpha}|K\rangle$.
In the notation of \Fig{fig:diag}, Lorentz invariance requires that this matrix element have the form
\begin{equation}
\label{hc}
\langle\pi|J^{\rm had}_{\alpha}|K\rangle=(P+p)_\alpha\:f_+(t)+(P-p)_\alpha\:f_-(t)
\end{equation}
where $t=(P-p)^2=(k+k')^2=M^2+m^2-2ME_\pi$ is the only $L$-invariant variable.
The form factors (FF) $f_+(t)$ and $f_-(t)$ account for the non-pointlike structure of the hadrons;
 the values of the FFs at $t=0$ differ from unity because of $SU(3)$ corrections, \ie, because
 pions and kaons have different structure.
The term containing $f_-$ is negligible for $K_{e3}$ decays, because
the coefficient $P-p=k+k'$, when acting on the leptonic current,
gives the lepton mass. The FF $f_-$ must be
retained for $K_{\mu3}$ decays. It is customary to introduce a scalar FF $f_0(t)$, such that \Eq{hc} becomes
  \begin{displaymath}
 \langle\pi|J^{\rm had}_{\alpha}|K\rangle=f(0)\left[(P+p)_\alpha\:\tilde f_+(t)+(P-p)_\alpha\:\left(\tilde f_0(t)\,{\Delta_{K\pi}\over t}-\tilde f_+(t)\,{\Delta_{K\pi}\over t}\right)\right],
 \end{displaymath}
 with $\Delta_{K\pi}=M^2-m^2$. Since the FFs $f_+$ and $f_0$ must have the same value at $t=0$,
 the term $f(0)$ has been factored out. The functions $\tilde f_+(t)$ and
 $\tilde f_0(t)$ are therefore both unity at $t=0$.
If the FFs are expanded in powers of $t$ up to $t^2$ as
\begin{equation}
\tilde f_{+,0}(t) = 1 + \lambda'_{+,0}~\frac{t}{m^2}+\frac{1}{2}\;\lambda''_{+,0}\,\left(\frac{t}{m^2}\right)^2
  \label{eq:ff2}
\end{equation}
 four parameters ($\lambda'_+$, $\lambda''_+$, $\lambda'_0$, and $\lambda''_0$)
 need to be determined from the decay spectrum in order to be able to compute the phase space
 integral that appears in the formula for the partial decay width.
 However, this parametrization of the FFs is problematic, because the values
 for the $\lambda$s obtained from fits to the experimental decay spectrum
 are strongly correlated, as discussed in the Appendix and in \Ref{pf}.
 In particular, the correlation between
 $\lambda'_0$ and $\lambda''_0$ is $-99.96\%$; that between
 $\lambda'_+$ and $\lambda''_+$ is $-97.6\%$. It is therefore impossible
 to obtain meaningful results using this parameterization.

Form factors can also by described by a pole form:
\begin{equation}
\fphat_{+,0} ={M_{V,S}^2\over M_{V,S}^2-t},
  \label{eq:pole}
\end{equation}
 which expands to $1+t/M_{V,S}^2+(t/M_{V,S}^2)^2$, neglecting powers of $t$
 greater than 2. It is not clear however what vector and scalar states
 should be used.

 Recent $K_{e3}$ measurements\cite{KTeV:ff,NA48:ff,KLOE:ff} show that the vector FF is
 dominated by the closest vector
 $(q\bar q)$ state with one strange and one light quark (or $K\pi$ resonance, in an older language).
 The pole-fit results are also consistent with predictions from a
 dispersive approach \cite{stern,sternV,pich1}.
 We will therefore use a parametrization for the vector FF based on a dispersion relation twice subtracted at $t=0$ \cite{sternV}:
\begin{equation}
\fphat = \exp\left[\frac{t}{m_\pi^2}\left( \Lambda_+ + H(t)\right)\right],
  \label{eq:sternv}
\end{equation}
where $H(t)$ is obtained using $K-\pi$ scattering data, and $\tilde{f}_+(0)=1$, $\tilde{f}'_+(0)= \Lambda_+/m_\pi^2$ .
A good approximation to \Eq{eq:sternv} is
\begin{equation}
\fphat =1+\lambda_+{t\over m^2}
+{\lambda_+^2+0.000584\over2}\left(t\over m^2\right)^2 
+{\lambda_+^3+3\times 0.000584\, \lambda_+ +0.0000299\over6}\left(t\over m^2\right)^3 .\nonumber
  \label{fplus}
\end{equation}
The errors on the constants 0.000584 and 0.0000299 in \Eq{fplus} are 0.00009 and 0.000002, respectively.
The pion spectrum in $K_{\mu3}$ decay has also been measured recently
 \cite{KTeV:ff,NA48:ff2,istra2}. As discussed in the Appendix, there is
 no sensitivity to $\lamopp$. All authors have fitted their data using a
 linear scalar FF:
\begin{equation}
\fzhat =1+\lambda_0 {t\over m^2}.
  \label{fzero}
\end{equation}
 Because of the strong correlation between $\lambda_0'$ and $\lambda_0''$,
 use of the linear rather than the quadratic parameterization gives a value
 for $\lambda_0$ that is greater than $\lambda_0'$ by an amount equal to
 about 3.5 times the value of $\lambda_0''$. To clarify this situation,
 it is necessary to obtain a form for \fzhat\ with $t$ and $t^2$ terms
 but with only one parameter.

 The Callan-Treiman relation\cite{ct} fixes the
 value of scalar FF at $t=\Delta_{K\pi}$ (the so-called Callan-Treiman point) to the
 ratio of the pseudoscalar
 decay constants $f_K/f_\pi$. This relation is slightly modified by SU(2)-breaking
 corrections\cite{dct}:
\begin{equation}
\tilde{f}_0(\Delta_{K\pi})=\frac{f_K}{f_\pi}\:{1\over f(0)}+\Delta_{CT},\quad \Delta_{CT}\simeq\pt-3.4,-3,
  \label{eq:ct}
\end{equation}
 A recent parametrization for the scalar FF \cite{stern}
 allows the constraint given by the Callan-Treiman relation to be exploited.
 It is a twice-subtracted representation of the FF at $t=\Delta_{K\pi}$ and $t=0$:
\begin{equation}
\tilde{f}_0(t)=\exp\left(\frac{t}{\Delta_{K\pi}} \log (C-G(t))\right)
  \label{eq:stern}
\end{equation}
 such that $C=\tilde{f}_0(\Delta_{K\pi})$ and $\tilde{f}_0(t)=1$.
 $G(t)$ is derived from $K\pi$ scattering data.
As suggested in \Ref{stern}, a good approximation to \Eq{eq:stern} is
\begin{equation}
\eqalign{\fzhat =1+\lambda_0 {t\over m^2}+{\lambda_0^2+0.000416\over 2}\left({t\over m^2}\right)^2
+{\lambda_0^3+3\times 0.000416\,\lambda_0+0.0000272\over 6}\left({t\over m^2}\right)^3 .\cr}
  \label{fzeroa}
\end{equation}
with $\log C= \lambda_0 \Delta_{K\pi}/m^2_\pi + 0.0398 \pm 0.0041$.
\Eq{fzeroa} is quite similar to the result in  \Ref{pich2}. The errors on the constants  0.000416 and 0.0000272 in \Eq{fzeroa} are 0.00005 and 0.000001, respectively.

At KLOE, the pion energy and therefore $t$ can be measured, since the
 \kl\ momentum is known at a \f\ factory. However, $\pi$-$\mu$ separation is
 very difficult at low energy. Attempts to distinguish pions and muons result
 in a loss of events of more than 50\% and introduce severe systematic
 uncertainties. We therefore use the neutrino spectrum, which can
 be obtained without $\pi$-$\mu$ identification.
\section{The KLOE detector}
\label{sec:kloedet}
The KLOE detector consists of a large cylindrical
drift chamber, surrounded by a
lead scintillating-fiber electromagnetic calorimeter.
  A superconducting coil around the calorimeter
provides a 0.52 T field.
The drift chamber \cite{KLOE:DC} is
4~m in diameter and 3.3~m long.
The momentum
resolution is $\sigma_{p_{\perp}}/p_{\perp}\approx 0.4\%$.
Two-track vertices are reconstructed with a spatial
resolution of $\sim$3~mm.

The calorimeter \cite{KLOE:EmC}
is divided into a barrel and two endcaps. It covers 98\% of the solid
angle.  Hits on cells nearby in time and space are grouped into
calorimeter clusters. The energy and time resolutions
are $\sigma_E/E = 5.7\%/\sqrt{E\ {\rm(GeV)}}$ and
$\sigma_T = 54\ {\rm ps}/\sqrt{E\ {\rm(GeV)}}\oplus100\ {\rm ps}$,
respectively.

The KLOE trigger \cite{KLOE:trig}
uses calorimeter and chamber information. For this
analysis, only calorimeter information is used.
Two energy deposits above threshold ($E>50$ MeV for the
barrel and  $E>150$ MeV for endcaps) are required.
Recognition and rejection of cosmic-ray events is also
performed at the trigger level. Events with two energy
deposits above a 30 MeV threshold in the outermost
calorimeter plane are rejected.

\section{Analysis}
\label{sec:analysis}
Candidate  \kl\ events are tagged by the presence of a \kspp\ decay.
The \kl\ tagging algorithm is fully described in Refs.~\citen{KLOE:brl}.
The \kl\ momentum, $p_{\kl}$, is obtained from the kinematics of the
$\phi\to\ks\kl$ decay, using the reconstructed \ks\ direction and the known value of $\pvec_{\phi}$.
The resolution is dominated by the beam-energy spread, and amounts to about
0.8 MeV/$c$. The position of the $\phi$\ production point, $\xvec_\phi$, is taken
as the point of closest approach of the \ks\ path to the beam line. The \kl\ line of flight
({\em tagging line}) is given by the \kl\ momentum,
$\pvec_{\kl} = \pvec_\phi-\pvec_{\ks}$ and the position of the production point,
$\xvec_\phi$.
 All tracks in the chamber, after removal of those from the \ks\
decay and their descendants, are extrapolated to their points of closest
approach to the tagging line.

For each track candidate, we evaluate the point of closest approach to the
 tagging line, \xc, and the distance of closest approach, \dc. The
momentum \pc\ of the track at \xc\ and the extrapolation length, \lc, are also
computed. Tracks satisfying $\dc< a \rxy + b$, with $a=0.03$ and $b=3$ cm, and
$-20<\lc<25$ cm are accepted as \kl\ decay products. \rxy\ is the distance of the
 vertex from the beam line. For each sign of charge, we chose the track with the
 smallest value of \dc\ as a \kl\ decay product, and from them we reconstruct the decay vertex.
Events are retained if the vertex is in the fiducial volume
$35<\rxy<150$ cm and $|z|<120$ cm.
The combined tracking and vertexing efficiency for \semil\ is about 54\%.
This value is determined from Monte Carlo (MC), corrected with the
ratio of data and MC efficiencies obtained from \kl\to\pic\po\,,$\pi e \nu$ control samples
\cite{KLOE:brl}.

Background from \kl\to\pic,\ \pic\po\ is easily removed by loose kinematic cuts.
The largest background is due to \klpen\ decays, possibly followed by early $\pi\to\mu \nu$
 decay in flight. For all candidate $K_{\mu3}$ events we compute
 $\rm{min}(\Delta_{\pi e},\Delta_{e \pi })$, the smaller value of $|\miss |$
 assuming the decay particles are $\pi e$ or $e \pi$.
 We retain events only if this variable is greater than 10 \mev.
 After the above kinematic cuts the efficiency for the signal is about 96\% and the purity is about 80\%.

A further cut  on the scatter plot of $\Delta_{\mu\pi}=E_{\rm miss}(\mu^{+},\pi^-)-p_{\rm miss}$ vs
 $\Delta_{\pi\mu}=E_{\rm miss}(\pi^{+},\mu^-)-p_{\rm miss}$ shown in the left panel of Fig. \ref{cut} for \kl\ \to\ $\pi \mu \nu$ and background events respectively, is applied.
\FIGURE{\figb pictures/12;6.4;\kern.4cm\figb pictures/mc-cut1;7;
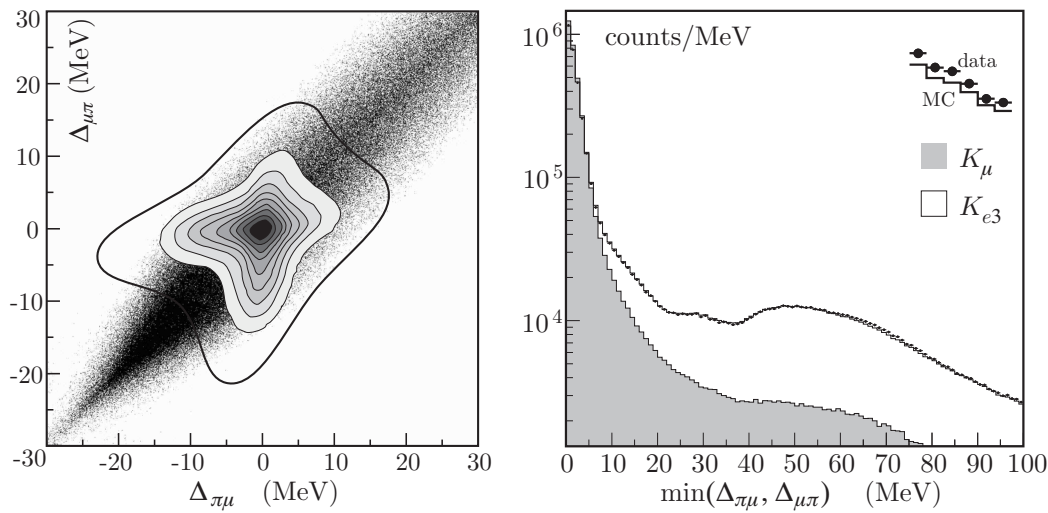
        \caption[examp of fig.]{Left: $\Delta_{\mu\pi}$ versus $\Delta_{\pi\mu}$ distribution from MC.  \kl\ \to $\pi\mu\nu$ (gray scale) and background (black points). The outsermost  contour shows the accepted region. Right: $min(\Delta_ {\pi\mu},\Delta_{\mu\pi})$ for data (black dots), MC (solid line), and MC signal (gray shaded histogram).}%
	\label{cut}}
The  right panel of  Fig. \ref{cut} shows the distribution of the lesser between $\Delta_{\pi\mu}$
and $\Delta_{\mu \pi}$ for data and MC.
After the kinematic cuts described above, the contamination, dominated by \kl\to $\pi e\nu$ decays is $\sim$4\%.

To further reduce  \kl\to $\pi e\nu$ background we use the  particle identification (PID) based on calorimeter information.
Tracks are required to be associated with EMC clusters. We define two  variables: \dTCA, the distance from the
extrapolated track entry point in the calorimeter to the cluster centroid and \dtTCA, the component of this distance in the
plane orthogonal to the track momentum at the calorimeter entry point.

We accept tracks with \dtTCA\ $<$ 30~cm.
The cluster efficiency is obtained from the MC, corrected
with the ratio of data and MC efficiencies obtained from control samples.
These samples, of 86\% and 99.5\% purity, are obtained from \semil\ and $K_{e3}$ events
selected by means of kinematics and independent calorimeter information.
The cluster efficiency correction versus $E_\nu$ is shown in Fig.~\ref{fig:tofcorr}.
\FIGURE{\figb pictures/cclu;5.5;
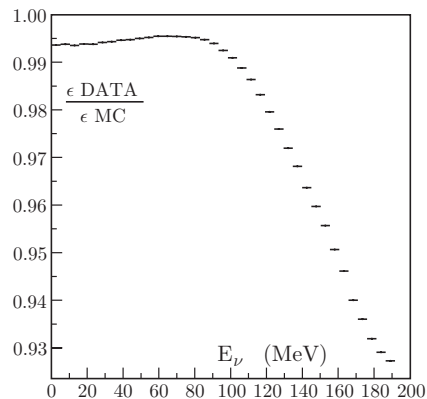
\caption{Cluster efficiency correction versus $E_\nu$.\label{fig:tofcorr}}}%
 For each \kl\ decay track with an associated cluster we define the variable:
 $\Delta t_i = t_{\rm cl}-t_{i} ,\ (i=\pi,\ e)$ in which $t_{\rm cl}$ is the
 cluster time and $t_{i}$ is the expected time of flight,
 evaluated using the corresponding mass. $t_{i}$ includes the time
 from the entry point to the cluster centroid~\cite{KLOE:kssemi}.
 We determine the $e^+e^-$ collision time, $t_0$, using the clusters from the \ks.

The mass assignment,
 $\pi e$ or $e\pi$,
is obtained by choosing the lesser of $|\Delta t_{\pi^+}-\Delta t_{e^-}|$ and
$|\Delta t_{\pi^-}-\Delta t_{e^+}|$.
After the mass assignment has been made, we consider the variable\vglue2mm
 $$R_{\rm TOF}=\left(\frac{\Delta t_{\pi} +\Delta t_e }{2\sigma_+}\right)^2
+\left(\frac{\Delta t_{\pi} -\Delta t_e }{ 2\sigma_-}\right)^2 $$
where $\sigma_+$ ($\sigma_-$) = $0.5 (0.4)$ ns  are the resolutions.
\FIGURE{\parbox{9cm}{\cl{\figb pictures/mc-cut2;7;}}
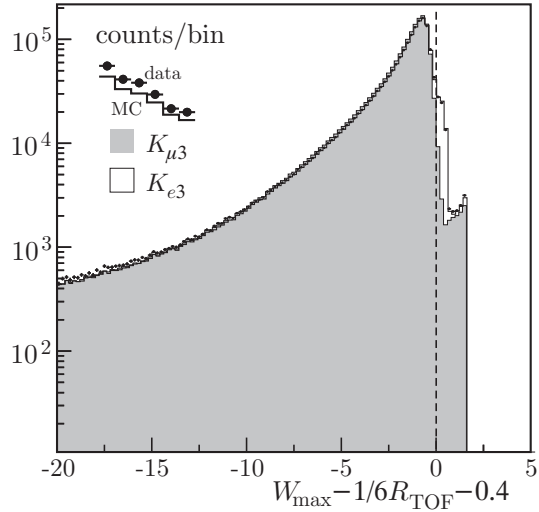
\caption{Distribution of $W_{\rm max}$ -1/6$R_{\rm TOF}-0.4$  for data (dots), MC (solid line) and MC signal (gray scale). The dashed line indicates the cut that we use.}
\label{fig:nntof}}
Additional information is provided by the energy deposition
 in the calorimeter and the cluster centroid depth. These quantities are input to a neural network (NN).
We retain events with $W_{\rm max}<(1/6)\,R_{\rm TOF}+0.4$, where $W_{\rm max}$  is the largest of the NN outputs for the two charge hypothesis. The distribution of $W_{\rm max}$ -1/6$R_{\rm TOF}-0.4$ for data and MC is shown in Fig.~\ref{fig:nntof}.
The resulting  purity of the sample is $\sim \,97.5\%$, almost uniform in range 16$<E_\nu<$181 MeV used for the fit. 

 The FF parameters are obtai\-ned by fitting the $E_{\nu}$ distribution of the selected
 events in the range $16<E_\nu<181$ MeV, subdivided in 32 equal width bins. The  bin size, ~5 MeV, is about 1.7  
 times the neutrino energy resolution.
 The value of $E_{\nu}$, \ie\ the missing momentum in the $K_L$ rest frame, is determined
 with a resolution of\break about 3 MeV almost independently on its value. The purity of the final sample used to extract the form factor parameters is illustrated in Fig.~\ref{fig:purity}.
\FIGURE{\figb pictures/purity;6;\caption{Purity versus $E_\nu$.}\label{fig:purity}}
 After subtracting the residual background as estimated from MC, we perform a $\chi^2$ fit to the data
 using the following expression for the expected number of events in each of the 32 bins:
\begin{equation}
\eqalign{N_i&=N_0\sum A_{ij}\times \Delta \Gamma_j (\mbox{\mathversion{bold}{$\lambda$}})\times\epsilon_{tot}(j)\cr
&\kern3cm\times~F_{FSR}(j),\cr} \label{fitfunc}
\end{equation}
where
$\Delta \Gamma_j(\mbox{\mathversion{bold}{$\lambda$}\mathversion{normal}})$ is the fraction of events expected for the parameter set defined by {\mathversion{bold}{$\lambda$}}
 in the $j^{th}$ bin,
 and $A_{ij}$ is the resolution smearing matrix. $F_{FSR}$ is the final state radiation correction.
It is evaluated using the MC  simulation, GEANFI ~\cite{KLOE:offline},
where radiative processes are simulated according the procedures described in \Ref{KLOE:gattip}.
FSR affects $E_\nu$-distribution mainly for high $E_\nu$ values, where the correction is about 2\%.
The free parameters in the fit are the FFs {\mathversion{bold}{$\lambda$}}.
 $N_0$ , the total number of signal events, is  fixed.

\section{Systematic uncertainties}

The systematic errors due to the evaluation of corrections, data-MC inconsistencies,
result stability, momentum mis-calibration, and background contamination are summarized
in \Tab{tabsyst}, for the case of a quadratic $\tilde f_{+}(t)$ and a linear  $\tilde f_{0}(t)$ .

\TABLE{
    \begin{tabular}{|c|c|c|c|}
      \hline
  Source &  $\delta\lamp\times 10^{3}$ & $\delta\lampp\times 10^{3}$ & $\delta\lamo\times 10^{3}$ \\
      \hline
      Tracking     &    1.60   &  0.47  & 0.86   \\
      Clustering   &    2.07   &  0.61  & 1.87   \\
      TOF + NN     &    2.23   &  1.16  & 1.45   \\
      p-scale &    1.10   &  0.71  & 0.81   \\
      p-resolution &    0.61   &  0.21  & 0.01   \\
      \hline
      Total        &    3.66   &  1.58  & 2.64   \\
      \hline
    \end{tabular}
    \caption{Summary of systematic uncertainties on \lamp, \lampp, \lamo.}
    \label{tabsyst}}%

The uncertainty on the tracking efficiency correction is dominated by sample
statistics and by the variation of the results observed using different criteria to
identify tracks from \kl\ decays. Its statistical error is taken into account in
the fit.
 We study the effect of differences in the resolution with which the variable
 \dc\ is reconstructed in data and in MC, and the possible bias introduced in the
 selection of the control sample, by varying the values of the
 cuts made on this variable when associating tracks to \kl\ vertexes.
 For each variation, corresponding to a maximal change of the tracking efficiency
 of about $\pm$10\%, we evaluate the complete tracking-efficiency correction
 and measure the slope parameters.
We observe  changes of 1.60$\times10^{-3}$,
  0.47$\times10^{-3}$, and 0.86$\times10^{-3}$
for  \lamp\, \lampp\ and \lamo\ , respectively.

 As for tracking, we evaluate the systematic uncertainties on the clustering efficiency
 corrections by checking stability of the result when the track-to-cluster association
 criteria are modified. The statistical uncertainty on the clustering efficiency corrections is taken into account in the fit.
 The most effective variable in the definition of track-to-cluster association
 is the transverse distance, \dtTCA. We vary the cut on \dtTCA\ in a wide range from
 15 cm to 100 cm,
 corresponding to a change in efficiency of about 19\%.
For each  value of the cut, we obtain the complete track extrapolation and clustering
efficiency correction and we use it to evaluate the slopes.
We observe   changes of  2.07$\times10^{-3}$, 0.61$\times10^{-3}$ and 1.87$\times10^{-3}$
for \lamp\,\lampp\  and \lamo, respectively.

 We study the uncertainties of the efficiency and of the background
 evaluation by studying the stability of the result with modified PID and
 kinematic cut values, corresponding to a variation of the cut efficiency from
  90\% to 95\%. This changes the background contamination
 from 1.5\% to 4.5\%.
 We observe  changes of
  2.23$\times10^{-3}$, 1.16$\times10^{-3}$ and 1.45$\times10^{-3}$
 for \lamp\,\lampp\  and \lamo, respectively.

We also consider the effects of uncertainties in the absolute momentum scale
and momentum resolution.
A momentum scale uncertainty of 0.1\% \cite{KLOE:offline}
corresponds to changes of $1.1\times10^{-3}$, $0.71\times10^{-3}$,
and $0.81\times10^{-3}$ for \lamp, \lampp, and \lamo, respectively.
 We investigate momentum resolution effects by changing the value
 of the resolution on $E_{\nu}$ by 0.1 MeV, an amount which characterizes
our knowledge of the momentum resolution, as described in \Ref{KLOE:ff}.
 We observe changes of
 $0.61\times10^{-3}$, $0.21\times10^{-3}$, and $0.01\times10^{-3}$
 for \lamp, \lampp, and \lamo, respectively.

\section{Results and interpretation}
 About 1.8 million \DKmuIII\ decays were accepted.
 We first fit the data using \Eqs{eq:ff2} \andEq{fzero} for the vector and scalar FFs.
The result of this fit is shown in \Fig{fig:fit_res}.
\FIGURE{\figb pictures/finalfit;7;
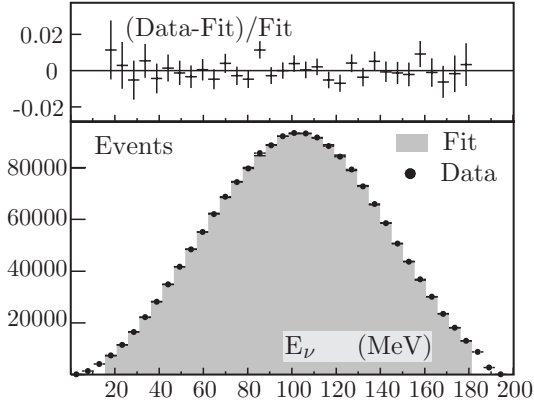
\caption{Residuals of the fit (top plot) and $E_\nu$ distribution for data events superimposed on the fit result (bottom plot).} \label{fig:fit_res}}
 We obtain:\\
\parbox{.45\textwidth}{\begin{eqnarray*}
  \lamp &=& (22.3\pm 9.8_{\rm{stat}}\pm 3.7_{\rm{syst}}) \times \rm{10^{-3}} \\
  \lampp &=& (4.8\pm 4.9_{\rm{stat}}\pm 1.6_{\rm{syst}}) \times \rm{10^{-3}} \\
  \lambda_0 &=& (9.1\pm 5.9_{\rm{stat}}\pm 2.6_{\rm{syst}}) \times \rm{10^{-3}}
\end{eqnarray*}}\\
\hglue.6cm\parbox{.35\textwidth}{\begin{displaymath}
\left(
\begin{array}{ccc}
 1 & -0.97 & 0.81 \\
   &     1 & -0.91 \\
   &       & 1     \\
\end{array}
\right)
\end{displaymath}}\\
with $\chi^2/{\rm dof}=19/29$, and correlation coefficients as given in the matrix.

Improved accuracy is obtained by combining the above results
 with those from our \DKeIII\ analysis\cite{KLOE:ff}:
\begin{eqnarray*}
  \lamp &=& (25.5\pm 1.5_{\rm{stat}}\pm 1.0_{\rm{syst}}) \times \rm{10^{-3}} \\
  \lampp &=& (1.4\pm 0.7_{\rm{stat}}\pm 0.4_{\rm{syst}}) \times \rm{10^{-3}} \\
\end{eqnarray*}
We then find:\\
\parbox{.5\textwidth}{\begin{eqnarray*}
  \lamp &=& (25.6\pm 1.5_{\rm{stat}}\pm 0.9_{\rm{syst}}) \times \rm{10^{-3}} \\
  \lampp &=& (1.5\pm 0.7_{\rm{stat}}\pm 0.4_{\rm{syst}}) \times \rm{10^{-3}} \\
  \lambda_0 &=& (15.4\pm 1.8_{\rm{stat}}\pm 1.3_{\rm{syst}}) \times \rm{10^{-3}}
\end{eqnarray*}}\kern1cm
\parbox{.35\textwidth}{\begin{displaymath}
\left(
\begin{array}{ccc}
 1 & -0.95 & 0.29 \\
   &     1 & -0.38 \\
   &       & 1     \\
\end{array}
\right)
\end{displaymath}}\\
with $\chi^2/{\rm dof}=2.3/2$ and the correlations given in the matrix on the right.

Finally, to take advantage of the recent parameterizations of the
FFs based on dispersive representations
(\Eqs{eq:sternv} \andEq{eq:stern}), we combine our results from
this analysis of \semil\ data with our previous result for $K_{e3}$.
We perform a fit to the values obtained for \lamp, \lampp, and \lamo\
that makes use of the total error matrix as described above, and
the constraints implied by \Eqs{fplus} \andEq{fzeroa}.
Thus, the vector and scalar FFs are each described by
a single parameter.
Dropping the ``$\;'\;$'' notations, we find
\begin{eqnarray*}
  \lambda_+ &=& (25.7\pm 0.4_{\rm{stat}}\pm 0.4_{\rm{syst}}\pm 0.2_{\rm {param}}) \times \rm{10^{-3}} \\
  \lambda_0 &=& (14.0\pm 1.6_{\rm{stat}}\pm 1.3_{\rm{syst}}\pm 0.2_{\rm {param}}) \times \rm{10^{-3}}
\label{final}
\end{eqnarray*}
with $\chi^2/{\rm dof}=2.6/3$ and a total correlation cofficient of
$-0.26$. The uncertainties arising from the choice of parameterization for
the vector and scalar FFs are $0.2\times 10^{-3}$ and
$0.1\times 10^{-3}$ using only \DKeIII\ decays and \DKmuIII\ decays,
respectively. These contributions to the uncertainty on
\lam\ and \lamo\ are explictly given in \Eq{final}.
We note that the use of \Eq{fzeroa} changes the value of the
phase space integral by only 0.04\% with respect to the result
obtained using a linear parameterization for \fzhat.

 Finally, from the Callan-Treiman relation we compute
$f(0) = 0.964\pm0.023$ using
 $f_K/f_\pi = 1.189\pm 0.007$ from \Ref{MILC}. Our value for $f(0)$
 is in agreement with the results of recent lattice calculations
\cite{f0:lattice}.

\section{Conclusions}
We have performed a new measurement of the  $\kl\to\pi\mu\nu$ FFs.
Our results are in acceptable agreement with the measurements from
KTeV \cite{KTeV:ff} and ISTRA+ \cite{istra2} and in disagreement with
those from NA48 \cite{NA48:ff2,NA48:ff}.
In particular, our result for the scalar FF parameter
$\lambda_0=0.0143\pm0.00203$ (\Eq{final}) accounts for the presence of
a $t^2$ term.
KTeV and ISTRA+ use a linear parameterization; as a consequence,
their values for $\lambda_0$ are systematically high by $\sim$0.003.
We also derive $f^{K^0}(0) = 0.964\pm0.023$. This value is in agreement with
the results of recent lattice calculations \cite{f0:lattice}.

\appendix
\section{Error estimates}
It is quite easy to estimate the ideal error in the measurements of a set of parameters {\bf p}=$(p_1,\:p_2,\ldots\:p_n)$ from fitting some distribution function to experimentally determined spectra.
Let $F(\mathbf{p},x)$ be a probability density function, PDF, where ${\mathbf p}$ is some parameter vector, which we want to determine and $x$ is a running variable, like $t$. The inverse of the covariance matrix for the maximum likelihood estimate of the parameters is given by \cite{hcra}:
$$({\bf G}^{-1})_{ij}=-{\partial^2 \ln L\over\partial p_i\partial p_j}$$
from which, for $N$ events, it trivially follows:
$$\left({\bf G}^{-1}\right)_{ij}=N\:\int{1\over F}\,{\partial F\over\partial p_i }\,{\partial F\over\partial p_j}\:\dif\upsilon,$$
with $\dif\upsilon$ the appropriate volume element.
We use in the following the above relation to estimate the errors on the FF parameters for one and two parameters expression of the FFs $\tilde f_+(t)$ and $\tilde f_0(t)$. The errors in any realistic experiment will be larger than our estimates, typically two to three times.  The above estimates are useful for  the understanding of the problems in the determination of the parameters in question.
\subsection{$K_{e3}$ decays}
For a quadratic FF, $\tilde f(t)=1+\lamp(t/m^2)+(\lampp/2)(t/m^2)^2$, the inverse of
the covariance matrix ${\bf G}_+^{-1}$, the covariance matrix ${\bf G}_+$ and the correlation matrix are:
$$N\,\pmatrix{5.937 & 13.867\cr\noalign{\vglue-4mm}\cr 13.867 & 36.2405\cr},\kern5mm
{1\over N}\,\pmatrix{1.258^2 & -0.606 \cr\noalign{\vglue-4mm}\cr -0.606 & 0.509^2},\kern5mm
\pmatrix{1 & -.945\cr\noalign{\vglue-4mm}\cr  & 1\cr}$$
The square root of the diagonal elements of ${\bf G}_+$ gives the errors, which for one million events are \dlp=0.00126, \dlpp=0.00051. The correlation is very close to \minus1, meaning that, because of statistical fluctuation of the bin counts, a fit will trade \lamp\ for \lampp\ and that the errors are enlarged.
A fit for a linear FF, $\tilde f(t)=1+\lamp(t/m^2)$ in fact gives \lamp=0.029 instead of 0.025 and an error smaller by \ab3:
$$\dlp=\sqrt{{\bf G}_+(1,1)}=0.0004.$$
A simple rule of thumb is that ignoring a $t^2$ term, increases \lamp\ by \ab3.5\x\lampp.
For $K_{e3}$ decays the presence of a $t^2$ term in the FF is firmly established. It is however not fully justified to fit for two parameters connected by the simple relation \lampp=2\x\lamp\up2. The authors of ref. \citen{stern,sternV} explicitly give an error for their estimate of the coefficient of the $t^2$ terms. The above discussion justifies the use of eq. \ref{fplus}. The errors obtained above compare reasonably with the errors quoted in \cite{KTeV:ff,NA48:ff,KLOE:ff}, when all experimental problems are taken into account.
\subsection{$K_{\mu3}$ decays}
The scalar FF only contributes to $K_{\mu3}$ decays. Dealing with these decays is much harder because: a) - the branching ratio is smaller, resulting in reduced statistics, b) - the $E_\pi$ or $t$ range in the decay is smaller, c) - it is in general harder to obtain an undistorted spectrum and d) - more parameters are necessary. This is quite well evidenced by the wide range of answers obtained by different experiments \cite{KTeV:ff,istra2,NA48:ff2}.
Assuming that both scalar and vector FF are given by quadratic polynomials as in eq. \ref{eq:ff2}, ordering the parameters as \lamop, \lamopp, \lamp\ and \lampp, the matrices ${\bf G}_{0\,\&\,+}^{-1}$ and ${\bf G}_{0\,\&\,+}$, are:
$$N\pmatrix{1.64&5.44&1.01&3.90\cr\noalign{\vglue-4mm}\cr
5.44&18.2&3.01&12.3\cr\noalign{\vglue-4mm}\cr
1.01&3.01&1.47&4.24\cr\noalign{\vglue-4mm}\cr
3.90&12.3&4.24&13.8\cr},\ {1\over N}\pmatrix{63.9^2&-1200&-923&197\cr\noalign{\vglue-4mm}\cr
-1200&18.8^2&272&-59\cr\noalign{\vglue-4mm}\cr
-923&272&14.8^2&-49\cr\noalign{\vglue-4mm}\cr
197&-59&-48&3.4^2 \cr}$$
and the correlations, ignoring the diagonal terms, are:
\begin{equation}
\label{4corr}
\pmatrix{-0.9996&-0.974&0.91\cr\noalign{\vglue-4mm}\cr
 &0.978&-0.919\cr\noalign{\vglue-4mm}\cr
 & &-0.976\cr}.
\end{equation}
All correlations are very close to \minus1. In particular the correlations between \lamop\ and \lamopp\ is \minus99.96\%, reflecting in vary large \dop\ and \dopp\ errors. We might ask what the error on \lamop\ and \lamopp\ might be if we had perfect knowledge of \lamp\ and \lampp. The inverse covariance matrix is give by the elements (1,1), (1,2), (2,1) and (2,2) of the ${\bf G}_{0\,\&\,+}^{-1}$ matrix above. The covariance matrix therefore is :
$${\bf G}_{0}(\lamop,\ \lamopp\ \hbox{for \lamp, \lampp\ known})={1\over N}\pmatrix{8.2^2&-20\cr\cr\noalign{\vglue -4mm}-20&2.4^2\cr}.$$
For one million events we have \dopp=0.0024, about 4\x\ the expected value of \lamopp. In other words \lamopp\ is likely to be never measurable. It is however a mistake to assume a scalar FF linear in $t$, because the coefficient of $t$ will absorb the coefficient of a $t^2$ term, again multiplied by \ab3.5. Thus a real value \lamop=0.014 is shifted by the fit to 0.017, having used eq. \ref{fzeroa}. Fitting the pion spectrum from 1 million $K_{\mu3}$ decays for \lamo, \lam\ with the FFs of eq. \ref{fzeroa} and \ref{fplus} gives the errors \dop\ab0.0096 and \dlp\ab0.00097. Combining with the result from a fit to 1 million $K_{e3}$ with the FF of eq. \ref{fplus} for which \dlp\ab0.00037 gives finally \dop\ab0.00075, \dlp\ab0.00034 and a \lamo-\lamp\ correlation of \minus31\%. Using the neutrino spectrum for $K_{\mu3}$ decays, the errors are only slightly larger: \dop\ab0.001, \dlp\ab0.00036. We hope to reach this accuracy with our entire data sample, \ab5\x\ the present one, and a better analysis which would allow using the pion spectrum.
\section*{Acknowledgements}
We would like to thank the authors of  \Ref{sternV} for providing us  with  the vector form factor parameterization. We thank the DA$\Phi$NE team for their efforts in maintaining low background running
conditions and their collaboration during all data-taking. We want to thank our technical staff:
G.F.Fortugno and F. Sborzacchi for their dedicated work to ensure an efficient operation
 of the KLOE Computing Center;
M.Anelli for his continous support to the gas system and the safety of the detector;
A. Balla, M. Gatta, G. Corradi and G. Papalino for the maintenance of the electronics;
M. Santoni, G. Paoluzzi and R. Rosellini for the general support to the detector;
C. Piscitelli for his help during major maintenance periods.
This work was supported in part
by FLAVIANET, 
by the German Federal Ministry of Education and Research (BMBF) contract 06-KA-957; 
by Graduiertenkolleg `H.E. Phys. and Part. Astrophys.' of Deutsche Forschungsgemeinschaft,
Contract No. GK 742; 
by INTAS, contracts 96-624, 99-37; 
and by the EU Integrated Infrastructure
Initiative HadronPhysics Project under contract number
RII3-CT-2004-506078.


\begin{thebibliography}{99}

\expandafter\ifx\csname url\endcsname\relax
  \def\url#1{\texttt{#1}}\fi
\expandafter\ifx\csname urlprefix\endcsname\relax\def\urlprefix{URL }\fi


\bibitem{KTeV:ff}
 T.~Alexopolous {\it et al.} [KTeV~Collaboration], Phys.\ Rev.\ D 70 (2004) 092007.

\bibitem{NA48:ff}
A.~Lai {\it et al.} [NA48~Collaboration], Phys.\ Lett.\ B 606 (2004) 1.

\bibitem{KLOE:ff}
F.~Ambrosino {\it et al.} [KLOE~Collaboration], Phys.\ Lett.\ B 636 (2006) 166.


\bibitem{pf} Paolo Franzini, Kaon Int. Conf. 2007, Opening remarks, PoS(KAON)002 (2007).

\bibitem{stern} V.~Bernard, M.~Oertel, E.~Passemar and J.~Stern, Phys.\ Lett.\ B {\bf 638} (2006) 480

\bibitem{sternV} V. Bernard, M. Oertel, E. Passmar and J. Stern, private communication. They compute a dispersion relation for ln $f_+$ twice subtracted at $t=0$, using $\pi K$ P-wave scattering data, as done in \cite{stern}. 

\bibitem{pich1} M.~Jamin, A.~Pich and J.~Portoles, Phys.\ Lett.\ B {\bf 640} (2006) 176

\bibitem{NA48:ff2}
A.~Lai {\it et al.} [NA48~Collaboration], Phys.\ Lett.\ B 647 (2007) 341.

\bibitem{istra2}
O. P. Yushchenko {\it et al.}, Phys.\ Lett.\ B 581 (2004) 31.

\bibitem{ct}
C.~G. Callan, S.~Treiman, Phys. Rev. Lett. 16 (1966) 153.

\bibitem{dct}
J.~Gasser, H.~Leutwyler, Nucl.\ Phys.\ B 250 (1985) 93.

\bibitem{pich2} M.~Jamin, J.~A.~Oller and A.~Pich, Phys.\ Rev.\ D {\bf 74} (2006) 074009


\bibitem{KLOE:DC}
M.~Adinolfi {\it et al.} [KLOE~Collaboration], Nucl. Instrum.\ Meth.\ A 488 (2002) 51.

\bibitem{KLOE:EmC}
M.~Adinolfi {\it et al.} [KLOE~Collaboration], Nucl.\ Instrum.\ Meth.\ A 482 (2002) 364.

\bibitem{KLOE:trig}
M.~Adinolfi {\it et al.} [KLOE~Collaboration], Nucl.\ Instrum.\ Meth.\ A 492 (2002) 134.

\bibitem{KLOE:brl}
F.~Ambrosino {\it et al.} [KLOE~Collaboration], Phys.\ Lett.\ B 632 (2006) 43;
M.~Antonelli, P.~Beltrame, M.~Dreucci, M.~Moulson, M.~Paultan, A.~Sibidanov,
  {Measurements of the Absolute Branching Ratios for Dominant \kl\ Decays, the
  \kl\ Lifetime, and $V_{us}$ with the Kloe Detector}, {KLOE Note 204} (2005).
  \newline{http://www.lnf.infn.it/kloe/pub/knote/kn204.ps.gz}

\bibitem{KLOE:kssemi}
F.~Ambrosino {\it et al.} [KLOE~Collaboration], Phys.\ Lett.\ B 636 (2006) 173, and
references therein.

\bibitem{KLOE:offline}
F.~Ambrosino {\it et al.} [KLOE~Collaboration], Nucl.\ Instrum.\ Meth.\ A 534 (2004) 403.

\bibitem{KLOE:gattip}
C.~Gatti, Eur.\ Phys.\ J.\ C, 45 (2006) 417.

\bibitem{MILC}
E. Follana et al. arXiv:0706.1726 [hep-lat] (2007).
\bibitem{f0:lattice}
  D.~Becirevic {\it et al.},
  Nucl.\ Phys.\  B {\bf 705}, 339 (2005);
  M.~Okamoto  [Fermilab Lattice, MILC and HPQCD Collaborations],
  Int.\ J.\ Mod.\ Phys.\  A {\bf 20}, 3469 (2005);
  N.~Tsutsui {\it et al.}  [JLQCD Collaboration],
  PoS {\bf LAT2005}, 357 (2006)
  [arXiv:hep-lat/0510068];
  C.~Dawson, T.~Izubuchi, T.~Kaneko, S.~Sasaki and A.~Soni,
  Phys.\ Rev.\  D {\bf 74}, 114502 (2006);
   D.~J.~Antonio {\it et al.},
  arXiv:hep-lat/0702026.

\bibitem{hcra} H. Cramer, Mathematical Methods of Statistics, Princeton University Press,
1946, proves that this is the smallest possible error.


\end{thebibliography}
\end{document}